# Unravelling the room temperature growth of two-dimensional h-BN nanosheets for multifunctional applications

Abhijit Biswas,*[a,g] Rishi Maiti,[b,g] Frank Lee,[c,g] Cecilia Y. Chen,[d] Tao Li,[e] Anand B. Puthirath,[a] Sathvik Ajay Iyengar,[a] Chenxi Li,[a] Xiang Zhang,[a] Harikishan Kannan,[a] Tia Gray,[a] Md Abid Shahriar Rahman Saadi,[a] Jacob Elkins,[a] A. Glen Birdwell,[f] Mahesh R. Neupane,[f] Pankaj B. Shah,[f] Dmitry A. Ruzmetov,[f] Tony G. Ivanov,[f] Robert Vajtai,[a] Yuji Zhao,[e] Alexander L. Gaeta,*[b,d] Manoj Tripathi,*[c] Alan Dalton[c] and Pulickel M. Ajayan*[a]

[a] Department of Materials Science and Nanoengineering, Rice University, Houston, Texas 77005, USA.

[b] Department of Applied Physics and Applied Mathematics, Columbia University, New York, 10027, USA.

[c] Department of Physics and Astronomy, University of Sussex, Brighton BN1 9RH, United Kingdom.

[d] Department of Electrical Engineering, Columbia University, New York, 10027, USA.

[e] Department of Electrical and Computer Engineering, Rice University, Houston, TX, 77005, USA.

[f] DEVCOM Army Research Laboratory, RF Devices and Circuits, Adelphi, Maryland 20783, USA.

[g] Abhijit Biswas, Rishi Maiti and Frank Lee equally contributed to this work

***Authors to whom correspondence should be addressed:** abhijit.biswas@rice.edu, m.tripathi@sussex.ac.uk, a.gaeta@columbia.edu, ajayan@rice.edu

**ABSTRACT**

Room temperature growth of two-dimensional van der Waals (2D-vdW) materials is indispensable for state-of-the-art nanotechnology. The low temperature growth supersedes the requirement of elevated growth temperature accompanied with high thermal budgets. Moreover, for electronic applications, low or room temperature growth reduces the possibility of intrinsic film-substrate interfacial thermal diffusion related deterioration of functional properties and consequent device performance. Here, we demonstrated the growth of ultrawide-bandgap boron nitride (BN) at *room temperature* by using the pulsed laser deposition (PLD) process and demonstrated various functionalities for potential applications. Comprehensive chemical, spectroscopic and microscopic characterization confirms the growth of ordered nanosheet-like hexagonal BN (h-BN). Functionally, nanosheets show hydrophobicity, high lubricity (low coefficient of friction), low refractive index within the visible to near-infrared wavelength range, and room temperature single-photon quantum emission. Our work unveils an important step that brings a plethora of applications potential for room temperature grown h-BN nanosheets as it can be feasible on any given substrate, thus creating a scenario for "*h-BN on demand*" at frugal thermal budget.

# Introduction

Two-dimensional van der Waals (2D-vdW) materials are astonishing in the nano-era, showing tremendous potential and technological relevance due to their unique atomic-scale growth, emergent functional properties, and application potentials.[1-4] In general, several top-down (chemical, and mechanical exfoliations) and bottom-up approaches (thin film growth by various physical/chemical vapor phase deposition techniques) have been employed to synthesize these 2D-vdW materials with atomic layer control.[5-7] Among these, the liquid exfoliation methods produce large-area 2D-vdW materials, however, the film quality is found to be very low (due to the presence of adsorbates and surfactants on the surface). Whereas vapor-phase thin film growth processes lead to smaller size crystals, but with high crystalline quality.[8] Due to this process-dependent variable quality, the growth of 2D-vdW materials thus remains to be an exceedingly intriguing research topic among the materials growth community. The formation, crystallinity, morphology, and consequent functional properties of 2D-vdW materials are highly sensitive to the thermodynamic and/or growth kinetics, adsorption of reaction species on a substrate surface, nucleation, and thus the resultant film growth and consequent properties.[9,10]

Among various procedures, for large-scale growth, intrinsic property evaluation, and nano-device fabrications, chemical vapor deposition (CVD) has widely been adopted to grow the 2D-vdW materials.[5,6] However, synthesizing 2D-vdW materials by CVD requires a very high-temperature of ~1000-1500 °C (sufficient supply of activation energy to adatoms to migrate to the energetically preferred locations onto the substrates during the growth); a rigorous synthesizing condition that limits device capabilities as elevated temperatures result in possible defect segregation and annihilation at the interfaces.[11-14] Therefore, researchers are giving efforts to synthesize 2D-vdW materials at relatively lower-temperatures, with the trade-off in crystalline domain size/quality, but generating novel application possibilities.[15-18] Theoretically, 2D material show anisotropy in growth and kinetics can be controlled by adjusting the balance between elements.[19] However, fundamentally, both thermodynamics (temperature) and kinetics play important roles, therefore temperature during the vapor-phase deposition has always been critical, which also affects nucleation density and grain size.[10,15] Therefore, the growth of 2D-vdW materials at lower or even at room temperature with a high-enough supply of kinetic energy directed towards reaction species might be sufficient to overcome the nucleation barrier, and if successful, would be very useful for several room-temperature applications. This would significantly enable the growth of 2D-vdW materials on any given substrate (with varying degrees of crystallinity, thickness, and uniformity and consequent emergent phenomena), creating a scenario for "*thin films growth of 2D-vdW materials on demand*", thereby ample investigations on properties and consequent applications.

To investigate this possibility of thin film *room-temperature* growth, we use hexagonal boron nitride (h-BN) as our model material. Among numerous 2D-vdW materials, ultra-wide bandgap h-BN (with a bandgap of ~5.9 eV) gets special attention for its excellent chemical inertness, high-thermal stability, and electronic, optical, and mechanical properties.[20,21] Electronically, h-BN is suitable as a gate-dielectric layer for the 2D-based field effect transistors (FET) devices.[22] High bandgap of h-BN makes it suitable for the fabrication of high-performance deep-ultraviolet photodetectors.[23] Moreover, weak inter-layer bonding in h-BN makes it a soft lubricant material and thus useful as a high-temperature corrosion resistance and antioxidation protective coatings for various industrial applications.[21] Given these advantages, we attempt to grow h-BN thin films at room temperature by using a highly energetic and thermally non-equilibrium thin film growth process, pulsed laser deposition (PLD). PLD possesses several advantages over other growth methods, as it is not in thermal equilibrium, directed onto the substrate, and the stoichiometry is preserved throughout the thin film deposition process, starting from a dense crystalline target of the desired material.[24] Several attempts have been implemented to grow h-BN films by PLD, however at elevated temperatures, showing mostly island-like growth.[25-30] The plasma formation during the ablation process of the target by pulsed UV-laser (photon energy of ~5 eV) contains radicals and ions (e.g. $B^+$, $N^+$, $N^*$, $N_2$, $N_2^+$, $N_2^*$ and $B^*N^+$) of the ablated species with kinetic energy ~10-100 eV (**Figs. 1a** and **1b**).[31]

We have grown h-BN at room temperature on various substrates: *c*-$Al_2O_3$ (0001), holey Cu-grid, and Si (100), and investigated their growth characteristics and several functional properties. Chemical, morphological, spectroscopic, and electron microscopy characterizations confirm the growth of ordered h-BN nano-sheets. We observed that films are hydrophobic, and exhibit a low-refractive index, and room temperature single-photon quantum emission. Moreover, the frictional characteristics reveal excellent lubrication properties with a low coefficient of friction. Thus, successful growth of ordered h-BN nano-sheets by PLD at room temperature might usher in the growth of h-BN on a variety of substrates, generating ample potential application under reduced thermal budgets, e.g. for flexible 2D-electronics,[32] coating layers for highly-corrosive materials, protecting layers for open-air degradable materials, precise photonic devices and for quantum information technology.

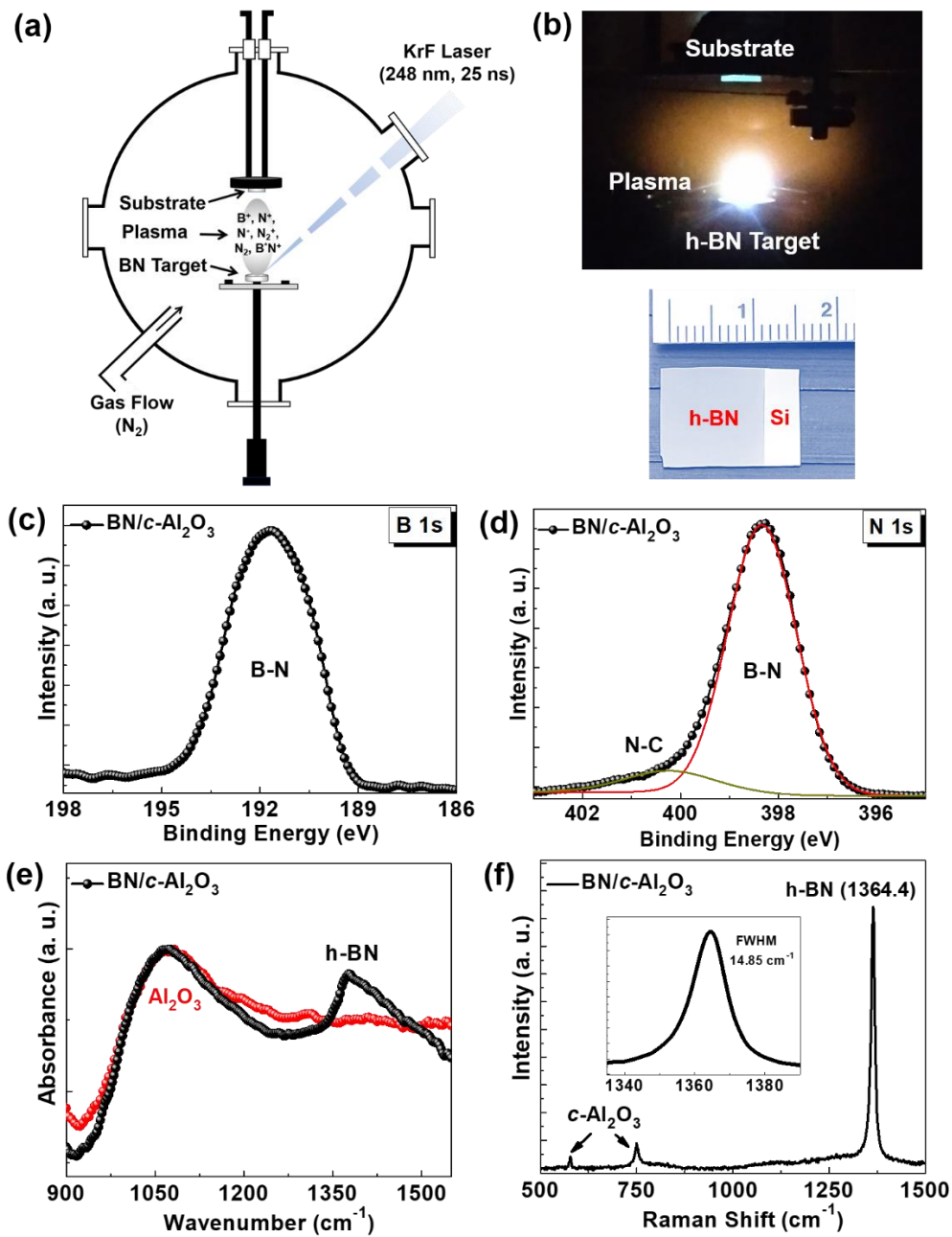

**Fig. 1 Pulsed laser deposition of h-BN and its spectroscopic characterizations**. (a) Schematic showing the thin film growth process by pulsed laser deposition. (b) Real-time plume generation (highly energetic) during the ablation of h-BN with ablated species deposited on a substrate kept

at room temperature. An optical image of a uniform (~1×1 cm$^2$) thick h-BN film (bottom). (c), (d) XPS elemental B 1s and N 1s core scans on *c*-Al$_2$O$_3$ confirm the presence of characteristics B-N bonds. (e), (f) FTIR and Raman spectra shows the peak at ~1364 cm$^{-1}$, further confirming the growth of h-BN at room temperature.

**Results and Discussion**

Structurally, to characterize the grown h-BN films we first performed conventional B 1s and N 1s core-level X-ray photoelectron spectroscopy (XPS) elemental scans. Characteristically, we observed the B-N bonding-related peaks in both the elemental B 1s (at ~190.6 eV) and N 1s (at ~398.5 eV) core level scans with the π-plasmon peaks (~9 eV apart from the main B-N peak), characteristics of h-BN (**Figs. 1c** and **1d** and Supplementary material **Figs. S1a** and **S1b**).[33] In N 1s scans, in addition to the B-N bonding peak, we observed a small shoulder of the N-C peak at ~400.5 eV (**Fig. 1d**), due to the ambient air exposure effect, related to adventitious carbon.[34] Through valence band spectra (VBS), we observed the characteristic band structure of the h-BN film (two distinct features with maxima at ~12 and ~20 eV, corresponding to the overlapping of σ and π-band as well as *s*-band, respectively) with valence band maxima (VBM) position at ~1.7 eV (Supplementary material **Fig. S1c**).[35,36] In order to confirm the phase, we then performed Fourier transformed infrared (FTIR) spectroscopy and Raman spectroscopy. FTIR shows a peak within 1350-1500 cm$^{-1}$, attributed to the in-plane B-N stretching vibrations (transverse optical mode) of the sp$^2$-bonded h-BN (**Fig. 1e**).[37] Moreover, in Raman spectra we observed Raman-active in-plane $E_{2g}$ phonon mode at ~1364 cm$^{-1}$ (**Fig. 1f**),[33] with a narrow full-width at half maxima (FWHM) of ~14.85 cm$^{-1}$, along with the peaks at ~577.4 and ~749.2 cm$^{-1}$, originating from the sapphire substrate. The Raman peak is slightly red shifted (~2 eV) to lower phonon frequencies compared to the bulk h-BN (~1366 eV), suggesting sheets are slightly tensile strained.[33] All these spectroscopic characterizations evince the room-temperature growth of h-BN.

Hereafter, we investigated the surface morphology of film by using atomic force microscopy (AFM). Interestingly, it shows 2D nano-sheet-like morphology (on *c*-Al$_2$O$_3$) (**Fig. 2a**). The lateral sizes of these h-BN sheets are ~200-300 nm and area coverage per 3×3 μm$^2$ is nearly ~70441.75 nm$^2$ (Supplementary material **Fig. S2**). Furthermore, to confirm the crystalline nature, we performed top-view high-resolution transmission electron microscopy (HRTEM) by growing h-BN films directly on holey Cu grid. For HRTEM, we have grown ultra-thin h-BN by providing only few numbers of laser shots (100 laser shots to be precise). As seen from the HRTEM (**Fig. 2b**), we also observed the nano-crystal-like feature. We captured several images at different magnification levels and obtained the diffraction patterns and *d* spacing. Remarkably, a closer inspection shows the appearance of clear ordered lattice fringes (**Fig. 2c**). We obtained *d* spacing which corresponds to the vdW gap between individual h-BN sheets ($d_{0002}$ = 0.33 nm) with a hexagonal diffraction pattern, although they exist in very weak intensity (**Fig. 2d** and Supplementary material **Fig. S3**).[20,21] TEM images, corresponding diffraction patterns and *d*-spacing also shows that along with the ordered hexagonal structure, some degree of amorphization (evident from the amorphous disk in the diffraction pattern) also exist in the nanosheets, atleast for the ultrathin h-BN grown on holy Cu-grid at room temperature.

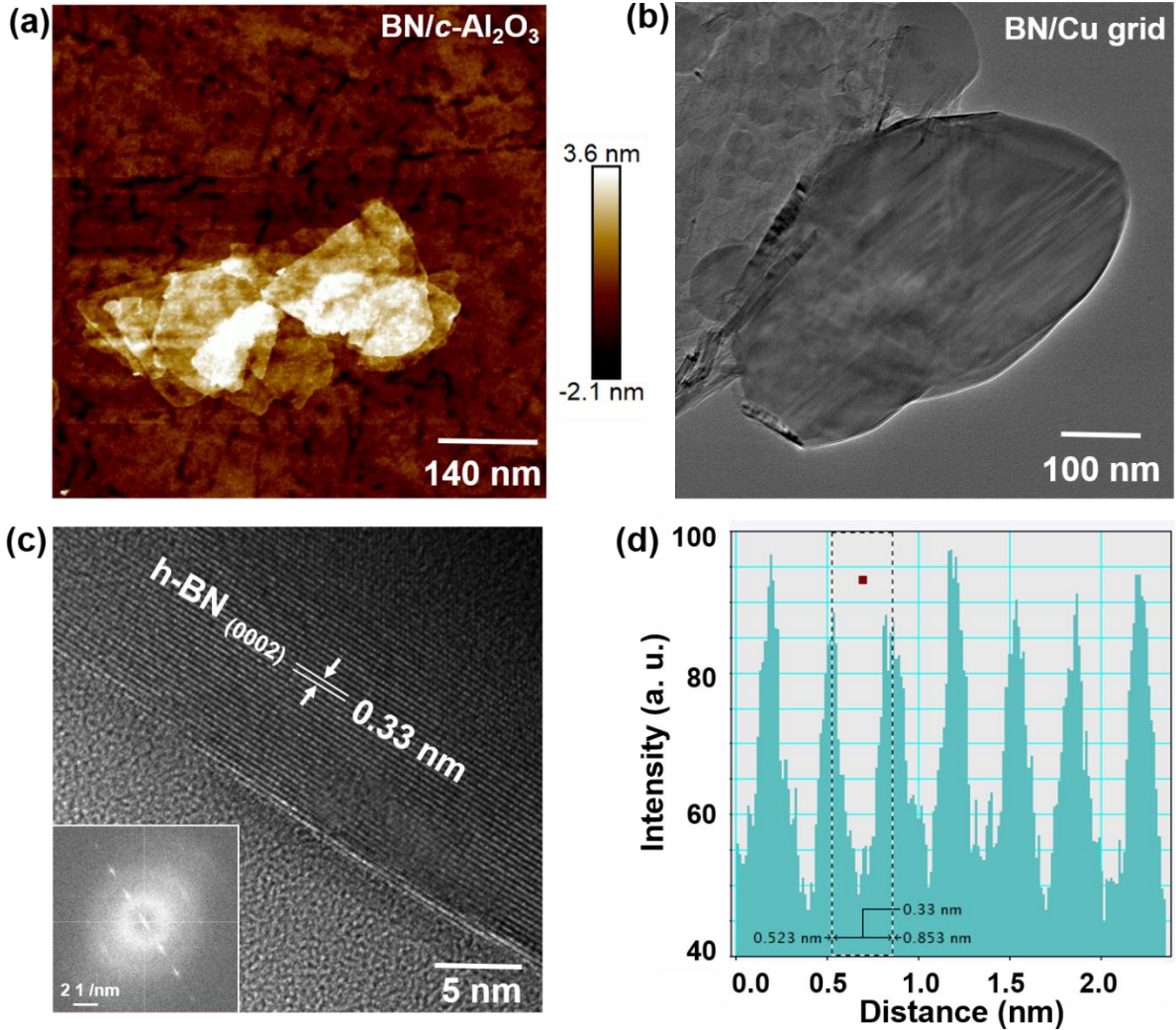

**Fig. 2 Microscopic characterizations of h-BN**. (a) AFM topography shows the nanosheets-like growth. (b), (c) High-resolution transmission electron microscopy (HRTEM) also shows the sheet-like structure with corresponding diffraction pattern (inset). (d) Inter-planar *d* spacing of ~0.33 nm, corresponding to the (0002) lattice plane of h-BN. For HRTEM, a few shots of BN was supplied directly on the holey Cu grid.

As mentioned earlier, the room temperature growth of highly ordered nanosheets is extremely interesting and cost-effective for scale-up, as it can be grown on potentially any given substrate (e.g. foils, flexible polymer surfaces), widening the horizon of h-BN growth for applications.[20] There have been attempts to grow h-BN at room temperature by RF sputtering and atomic layer depositions (ALD), but have come up with an amorphous/turbostratic phase.[38,39] In the case of PLD, which is most often used for metal-oxides thin film depositions, it is well known that high-temperature (≥700 °C) is needed for epitaxial growth (depending on the thermodynamics).[24] However, as reported by Elhamid *et al.*, well-defined graphene can be grown at room temperature by PLD on a bi-metallic Ni-Cu substrate, where laser power played a crucial role, favoring nucleation activation energy which promotes graphene formation.[40] Rasic *et al.* grew metallic TiN thin films on sapphire at room-temperature.[41] Kakehi *et al.* reported the epitaxial growth of NiO (111) thin films by PLD at room-temperature.[42] Ma *et al.* grew ZnO films on glass at room temperature by PLD.[43] Fundamentally, adsorption energy is one order higher than the diffusion energy barrier and at lower temperature, the adsorption rate exceeds the desorption rate.[15] Therefore, small-fractal shaped domain growth has been observed for vapor-phase synthesis for 2D materials at high flux and at reduced temperature.[15] Very recently, people have also started putting efforts to grow h-BN at various temperatures by PLD.[44] Hence, from growth perspective, an energetic bombardment of particles during the growth process is necessary to synthesize h-BN film at lower/room temperature, with the ability for the adatoms to overcome the substrate diffusion energy barrier. Ideally, thin film synthesis process involves: (1) evaporation and transport of ablated species to substrate, (2) migration of ablated species onto the substrate, and (3) nucleation and crystal growth.[24] For h-BN growth, the laser ablated species are atomic neutral and ionized species including $B^+$, $B^*$, $N^+$, $N^*$, and molecular species including $N_2^*$, $N_2^+$, and $B^*N^+$.[31] These species arrives concurrently on the substrate with a high kinetic energy of ~10-100 eV, and thereby forming h-BN layers (nanosheets for our case) within microseconds (~2 μS) onto the substrate, even at room temperature. The importance of the substrate temperature is to supply additional energy for these species to mobilize onto the surface, coalescence, make bonding with the substrates, thus promoting better nucleation, and consequent crystallinity. We have also performed the FESEM and grazing incident angle x-ray diffraction (GIXRD) of the h-BN target, before and after the laser ablation (Supplementary material **Fig. S4**). Both cases, FESEM show similar morphology with the typical sheet sizes of several µm, which is much larger than the lateral sizes for the room temperature grown h-BN nanosheets, as shown in AFM (**Fig. 3b**). The GIXRD shows only the reduction in (0002) h-BN peak intensity after the laser ablation, apparent as the target surface becomes rougher after the ablation. Therefore, considering PLD is a highly energetic deposition process and in addition BN contains light elements, thus synergistically, created vapor pressure and kinetic energy of the laser-ablated species in the plasma form possibly sufficient to migrate to the substrate and nucleate onto it (within a few µS) and forms the films even at room temperature, affecting the nucleation density and grain size,[15,24] All these observations reduces the possibility that growth happens due to the sputtering of nano-BN particles from the target, rather it follows the typical thin film growth process, at least for the present employed growth protocols.

To realize the intrinsic functional properties of the h-BN film, first, we investigate the wetting characteristics by the contact angle (CA) measurement, as hydrophobic surfaces are known as effective protective layers in harsh environments with extreme temperatures, and h-BN provides strong resistance to chemical attack and high-temperature stability.[20] As observed, h-BN tends toward minimizing the contact surface with water. Films exhibit hydrophobic behavior which is fairly increased by ~5%, (CA: 55°±1 to 60°±1) (**Fig. 3a**), consistent with the fact of the polar character of the boron-nitrogen bond ($–N^-B^+–$), as the wettability is strongly influenced by the vdW interactions.[45] The measured CA is also in agreement with the reported values ranging between ~50-67°.[45] Fundamentally, the wetting characteristics of a particular material also depend on the surface roughness and are inherent to the wetting phenomena.[46,47] However, considering that the surface roughness of grown h-BN film is of few nm (~2-3 nm); the effect of roughness on the CA may be insignificant here.

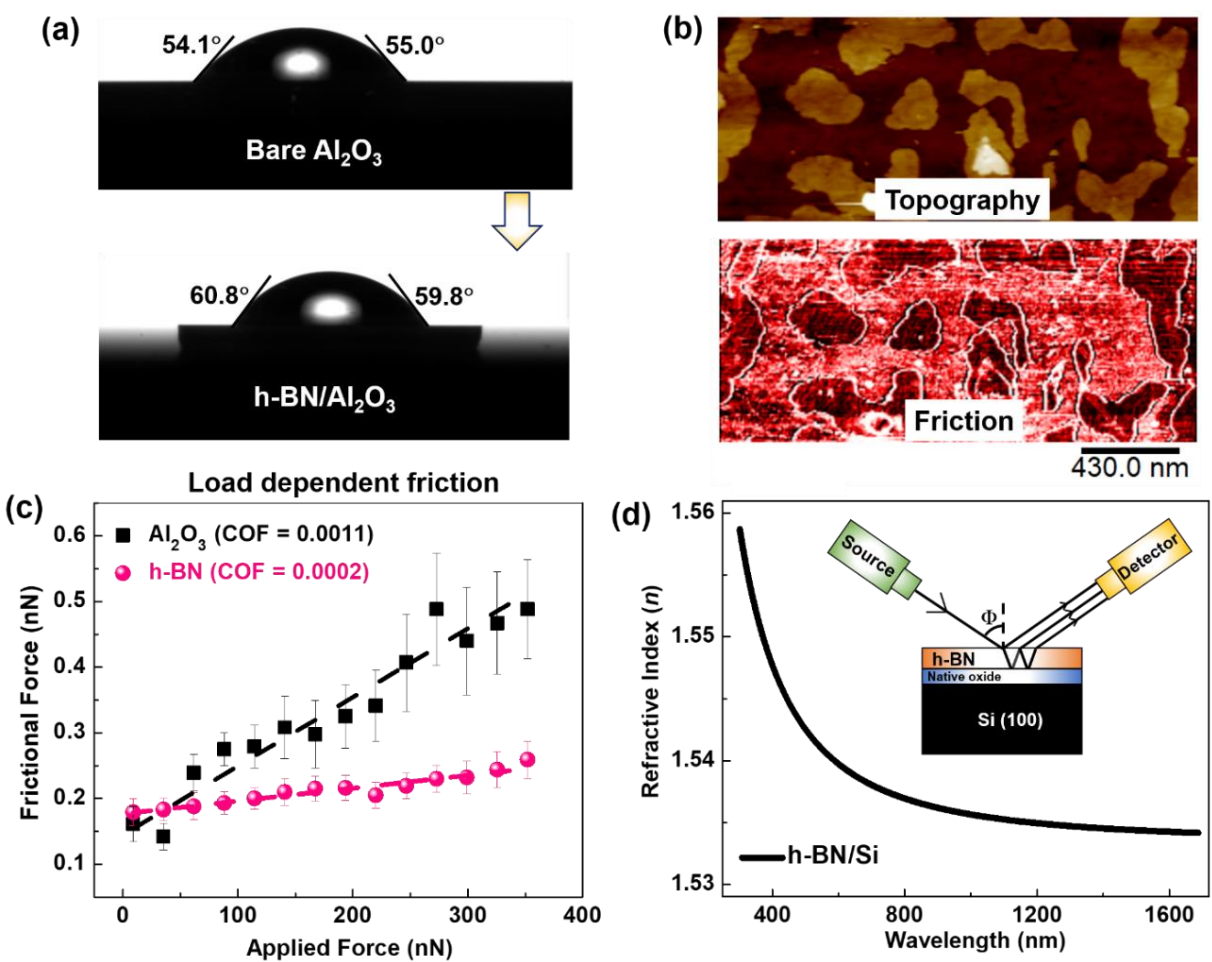

**Fig. 3 Functional properties of h-BN**. (a) Contact angle measurement on the h-BN surface, showing the water repelling nature. (b) Topography (upper panel) and the corresponding frictional force (lower panel) map of h-BN/$c$-Al$_2$O$_3$. Darker colors in the frictional force map at the h-BN suggest lower friction. (c) The measured frictional force with different loading. The slope of the fitted line represents the coefficient of friction (COF) of the material. (d) The refractive index ($n$) of the h-BN film (grown on Si substrate). The inset shows the schematic of the measurement.

The effect of hydrophobicity and inertness of h-BN nanosheets are further investigated through force-distance (F-D) spectroscopy measurements and its associated line profile using AFM probe covered with diamond-like carbon (DLC) (Supplementary material **Figs. S5** and **S6**). In the F-D measurement, the AFM probe temporarily brings a closer "repulsive regime" to the h-BN nanosheets for short durations (sub-seconds) and then moves away vertically from the surface. It requires the work against adhesion force that brings the cantilever away from the surface (i.e. h-BN nanosheet and $c$-Al$_2$O$_3$), referred as a "pull-out" force. The higher value of pull-out force indicates vital interaction with the substrate towards tip apex and vice versa. The h-BN nanosheets show lesser pull-out force values towards DLC tip as compared to the host substrate corroborating CA measurements.

At the submicron scale, the intrinsic roughness and moderately hydrophobic surface influence the tribological response. Thus, the frictional characteristics of h-BN (on $c$-Al$_2$O$_3$) is examined by lateral force microscopy where the topography and the corresponding frictional force map are measured at the same acquisition (**Fig. 3b**). The presence of h-BN layers anticipates lower friction force values than the surrounded $c$-Al$_2$O$_3$ substrate. However, the edges and step-edges lead to the highest friction values due to the higher susceptibility of edge atoms towards the slider (tip apex) which is peculiar to the family of 2D-vdW materials.[48,49] The tendency of the lubrication is further monitored where the probe sweeps the same region of h-BN and $c$-Al$_2$O$_3$ with increasing applied normal force ranging from 10 nN to 350 nN (**Fig. 3c**). We observed a linear response of increase in friction force as the function of normal forces, where the trend of load-dependent friction is higher for the alumina substrate as compared to the h-BN, quantified through linear fit resulting in the coefficient of friction (COF). The COF values of h-BN and $c$-Al$_2$O$_3$ are obtained as ~0.0002 and ~0.0011, respectively. The COF value depends on the nature of the material used and the local contact area between the tip apex and the substrate. Considering the local contact area of the tip apex is similar while rubbing between h-BN and alumina surface in a single scan at fixed load conditions. It is the nature of the material dominating the higher COF values for alumina. The additional factor influencing the frictional behavior in 2D materials is interfacial interaction with the host surface. It is observed that $c$-Al$_2$O$_3$ plane is relatively reactive w.r.t. the other planes (e.g. $a$-Al$_2$O$_3$), to establish a stronger interfacial bond with h-BN.[50] Weak vdW interaction (e.g. h-BN-silica, graphene-silica) leads to higher friction force of the atomically thin sheet owing to the puckering effect "i.e. out-off plane deformation that resists sliding" than the thick layers.[49] While, stronger interfacial interaction (i.e. graphene-mica, and graphene-Ni (111)) does not show the puckering effect, revealing that the frictional force response is independent of thickness.[51,52] We did not observe thickness dependent frictional contrast in our friction map, indicating stronger interaction between h-BN and $c$-Al$_2$O$_3$ (Supplementary material **Fig. S2**). These results are in good agreement with conventionally fabricated h-BN nanosheets and other 2D materials.[52-54] Overall, the observed hydrophobic nature and lubricity of h-BN nanosheets would be useful for the coating industry.

Furthermore, we measured the refractive index (*n*) of an h-BN film grown on Si substrate (see Supplementary Information **Fig. S7** for the details characterizations of the film), which was found to be ~1.53-1.55 (**Fig. 3d**) within the visible and near-infrared spectral range. We used variable angle spectroscopic ellipsometry (VASE) to measure the refractive index (*n*). The schematic of the measurement is depicted in the inset of **Fig. 3d**. Considering the nanosheet-like morphology of h-BN, thus we fitted the ellipsometry data by applying the Bruggeman effective-media approximation (EMA) method that considers the film as h-BN with defects (See method section and Supplementary material **Fig. S8**). Interestingly, the h-BN with added voids into the fitting model provides lower *n* than the perfect single crystalline layer h-BN fitting (*n* ~1.8). The lower *n* is adequate as stoichiometry, porosity, roughness affects the value of *n*.[55,56] Interestingly, the lower *n* is helpful for the precise designing of photonic devices operating in visible and near-infra-red wavelengths range.[55]

Since nanosheet-like h-BN can host defects such as point defect including boron-vacancy ($V_B$) and/or nitrogen-vacancy ($V_N$), and nitrogen-vacancy and boron replacing nitrogen ($V_NN_B$), and also carbon impurities, thus it may show single photon emission (SPE) at room temperature, attributed to these localized defects states inside the bandgap.[57-61] SPE in h-BN is extremely important for the next-generation quantum computing and information-processing technologies, considering its excellent chemical and thermal robustness.[62,63] **Figure 4a** shows confocal photoluminescence (PL) map (~50×50 μm$^2$) of h-BN nano-sheets showing much brighter emission spots (yellow) above the background.

**Figure 4b** shows the corresponding PL emission spectrum under continuous wave non-resonant excitation (λ =532 nm). The emitter has a sharp zero-phonon line (ZPL) at ~577 nm with an asymmetric shape (FWHM ~2 nm), extracted from a single Lorentzian fit. A weak optical phonon replica is visible around ~630 nm, red-shifted by ≥165 nm from the ZPL peak, related to phonon sidebands (PSB).[64] It should be noted that the small peak at ~580 nm is most likely a ZPL of another weak emitter within the excitation spot. In order to understand the extent of the electron-phonon coupling we calculate the Debye-Waller factor (the ratio between ZPL intensity to the total emission) is found to be ~ 0.70.

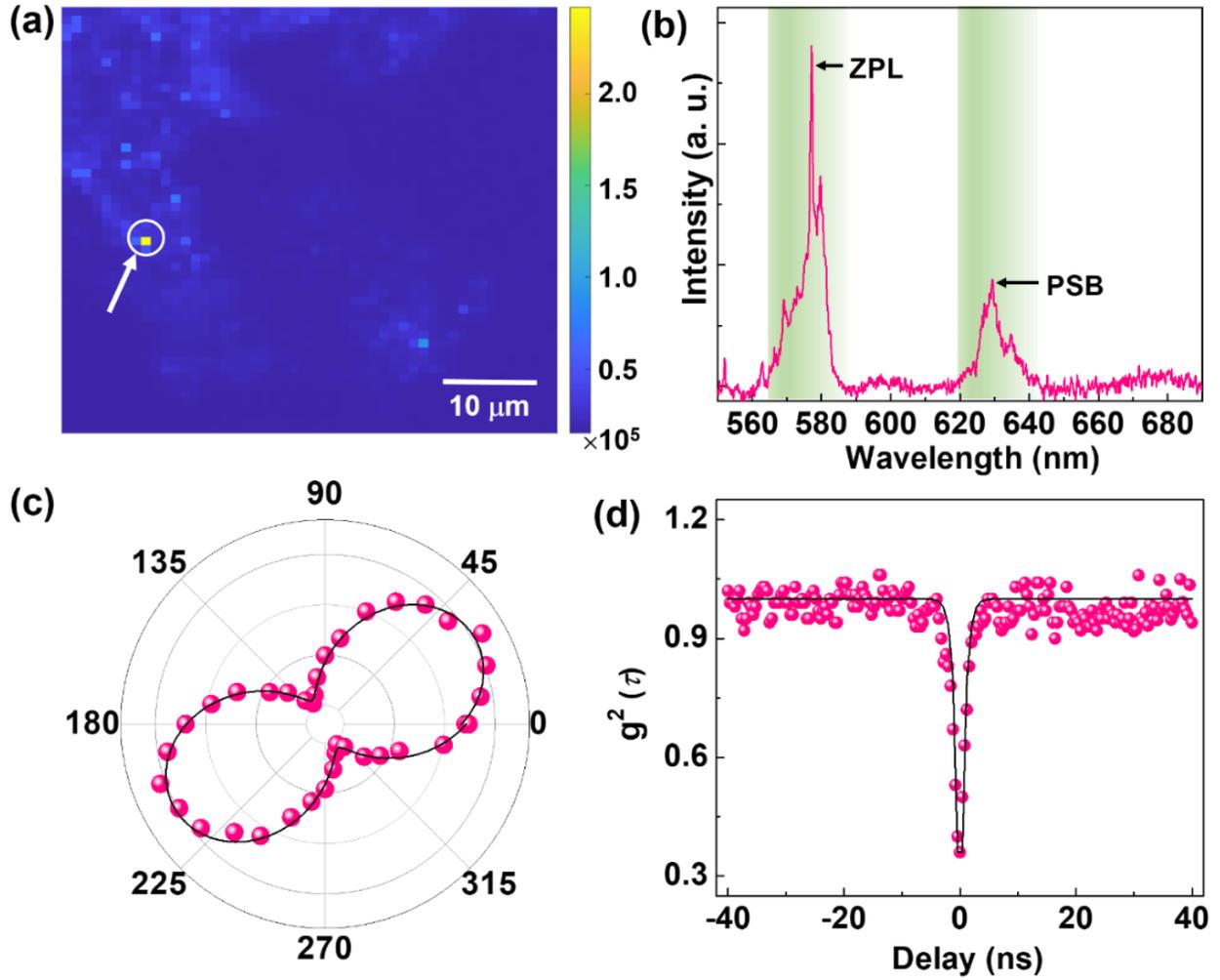

**Fig. 4 Room-temperature photo-physical properties of h-BN nanosheets**. (a) Confocal photoluminescence map (~50×50 μm$^2$) corresponds to the ultra-bright (yellow spots) defect-assisted emission from h-BN nano-sheets grown on sapphire. (b) Photoluminescence spectrum under 532-nm laser excitation and a 550-nm long-pass filter at room temperature. (c) The emission (red spheres) polarization curve. The red curve is fitted using a $\cos^2\theta$ function indicating an atomic defect with the in-plane dipole moment. (d) Second-order autocorrelation measurement $g^{(2)}(\tau)$ of the defect with a dip at $g^{(2)}(0) \approx 0.36$, suggesting the antibunching nature of light.

In order to understand the polarization behavior of the localized emission, we performed emission polarization measurements. **Figure 4c** corresponds to the angular diagram of emission, which reflects a well-defined dipole-like character for the single isolated defect. The corresponding fits are obtained using a $\cos^2(\theta)$ fitting function, which yields an emission polarization visibility of ~75%, suggesting a single dipole transition linearly oriented along the basal plane of the h-BN nanocrystals. **Figure 4d** displays the second-order correlation function $g^2(\tau)$ using two Si APDs, mounted in a Hanbury-Brown and Twiss (HBT) configuration to establish the quantum nature of the emitted light from the defects of h-BN nanocrystals. The $g^2(0)$ is found to be ~0.36, well below of 0.5 (without background correction) at zero delay time (**Fig. 4d**), which clearly indicates the anti-bunching of the emission.[62] The experimental $g^2(\tau)$ data can be fitted well with the three-level model: $g^2(\tau) = 1-(1+a)e^{-|\tau|/\tau_1} +ae^{-|\tau|/\tau_2}$, where a is a fitting parameter, $\tau_1$ and $\tau_2$ are the life-times of the excited and metastable states, respectively. Moreover, the radiative lifetime $\tau_1$ ~3 ns is consistent with that of the previous report,[64] however the non-radiative lifetime $\tau_2$ is much longer. These observations clearly show the room temperature stable SPE in h-BN. Our results indicate this stable and bright SPE observed in the visible range at room temperature potentially could solve the long-term issues of the on-chip integration of deterministic SPEs over large scales. Our approach of growing h-BN nanosheets even at room temperature depicts the ability to directly grow 2D layers onto designed electronic/photonic structures, which can be very useful for integrated circuits applications since it avoids any chemical contaminants or material deformation that often occurs with the wet or dry transfer.

In view of device applications, h-BN nanosheets show excellent promises. 2D h-BN has emerged as a suitable material for the fabrication of cost-effective deep-ultraviolet (UV) photodetectors.[65] Recently, it was shown that h-BN nanosheets could be useful as a saturable absorber for mode-locked lasers.[66] h-BN can act as a surface passivation layer for reducing the electrical loss of 2D-vdW based solar cells.[67] h-BN could also be useful as a tunnel barrier in 2D-vdW heterostructures and devices for information science and nanotechnology.[68] Considering that h-BN could be grown on Si at room temperature, this would solve the thermal stability related issue of Si substrtae while the high-temperature growth of h-BN on it, thus would be advantageous for the design of Si-based photonics integrated.[69] For energy harvesting, h-BN nanosheets based devices are also useful to develop a high-performance hybrid piezo/triboelectric nanogenerator (PTEG) by incorporating it into the polydimethylsiloxane (PDMS).[70] Therefore, room temperature in-situ grown chemical contaminant free h-BN nanosheets would be extremely beneficial for the next-generation cost-effective clean optoelectronic and energy harvesting applications.

**Conclusion**

In summary, we have shown that by using the highly energetic pulsed laser deposition (PLD) vapor phase thin film growth process, ordered nanosheets of 2D-vdW h-BN could be grown on desirable substrates, and remarkably, even at room temperature. This is attributed with the ability for the adatoms to overcome the substrate diffusion energy barrier, because of the interplay between the thermodynamics and the high kinetics of PLD process. We used extensive characterizations including chemical, spectroscopic, microscopic, tribological and optical tools to confirm the growth of ordered 2D-vdW h-BN nanosheets with the exhibition of various properties. Functionally, h-BN nanosheets show improved-hydrophobicity, low refractive index in the visible wavelength range. Furthermore, it shows excellent lubricity with a reduction in the friction force coefficient than that of the subsurface. Astonishingly, films also exhibit room temperature single-photon emission, an important building blocks of the optical quantum nanotechnologies. Therefore, considering the humongous application-worthiness, our observation of successful growth of 2D-vdW h-BN nanosheets at room temperature, might usher its growth on any substrates (also might show the pathway of growth other 2D-vdW materials at room temperature), generating ample potential applications under reduced thermal budgets, e.g. flexible 2D-electronics, coating layers for highly corrosive and degradable materials, precise photonic devices and for quantum information technology, thus potentially creating a scenario of "*h-BN on demand*" under a frugal thermal budget, indispensable for nanotechnology.

## Methods

**Thin film growth (Pulsed Laser Deposition):**

h-BN nanosheets were grown by PLD (KrF excimer laser with operating wavelength of 248 nm, and pulse width of 25 ns). Films were grown at room-temperature by using the following deposition conditions: laser fluency ~2.2 J/cm$^2$ (laser energy ~230 mJ, spot size ~ 7 mm ×1.5 mm), repetition rate 5 Hz, high-purity (5N) nitrogen gas partial pressure ($P_{N2}$) ~100 mTorr (flow rate ~73 sccm), and target to substrate distance ~50 mm. We used commercially available one-inch dense high-purity (99.9% metal basis) polycrystalline h-BN target for the ablation (American Element). The chamber base pressure was 5×10$^{-9}$ Torr. For substrates, we used hexagonal $c$-Al$_2$O$_3$ (0001), Si (100), and holey Cu grid as substrates with 2000 laser pulses (on Al$_2$O$_3$), 5000 laser pulses (on Si) and 100 laser pulses (on holey Cu grid) was given to the target, respectively. Substrates were purchased from MSE suppliers, USA.

**Chemical and microscopic characterizations (XPS, VBS, FTIR, AFM, Raman, GIXRD, FESEM, and HRTEM)**

X-ray photoelectron spectroscopy was performed with a PHI Quantera SXM scanning X-ray microprobe with 1486.6 eV monochromatic Al K$_\alpha$ X-ray source. High-resolution scans were recorded at 26 eV pass energy. The valence band spectrums were recoreded at 69 eV pass energy. XPS curve fitting was also performed using Shirley baseline by using the Multipak XPS software. FTIR was obtained by using the Nicolet 380 FTIR spectrometer (4000-400 cm$^{-1}$ region), equipped with a single-crystal diamond window (resolution 8 cm$^{-1}$). Park NX20 AFM was used to obtain surface topography, operating in tapping mode using Al-coated Multi75Al cantilevers. Raman spectroscopy was performed using a Renishaw inVia confocal microscope, with a 532 nm (50× resolution, 10% power) laser as the excitation source. Grazing incident angle X-ray diffraction (GIXRD) (with incident angle of 0.1°) was obtained with Rigaku SmartLab X-ray diffractometer, by using a monochromatic Cu Kα radiation source (λ = 1.5406Å ). The h-BN target surface topography was performed by field emission scanning electron microscope (FESEM) (FEI Quanta 400 ESEM FEG). We sputtered ~10 nm gold on h-BN for the FESEM measurements. For HRTEM, the ultra-thin film was directly grown on holy Cu-gird and the images were recorded using Titan Themis operating at 300 kV and the images were obtained in different magnification level.

**Adhesion and friction force characterizations**

Atomic force microscopy (from Bruker Ltd.) was operated in Bruker ScanAsyst using peak force tapping mode (Bruker's proprietary) with an applied normal force of 1nN. All the AFM operations were carried out at room temperature under ~35-40% of relative humidity. Several AFM tip materials (i.e. cantilever) have been used in the measurement to investigate the surface chemistry (adhesion force) between h-BN, substrate and tip apex. Silicon nitride cantilever (model: Scanasyst-air (Bruker), stiffness ~0.5± 0.15 N/m) was used for the topography measurements. Diamond-like carbon coated tip (model: DT-CONTR (Nanosensors), stiffness ~40± 2 N/m) used for Force-distance spectroscopy and mechanical mapping. The friction force (nN) was measured in the mode of "lateral force microscopy", here silicon cantilever (model: CSC38 (Micromash) of stiffness (0.14 ±0.05) N/m by Sader's method has been used for the investigations.[71] The average values of lateral force in trace and retrace direction are measured as half times of trace minus retrace (TMR). This procedure is useful to mitigate the topographical effect in the friction measurements. The tip radii are evaluated through the blind tip reconstruction technique using a grating sample as described in previous studies.[72]

**Optical characterizations (Refractive index and single-photon emission)**

Variable angle spectroscopic ellipsometry (VASE) (M-2000 Ellipsometer by J. A. Woollam Company) was applied to measure the refractive index (RI). The RI of a bare silicon substrate with a thin native oxide layer on the top is measured first as a reference, followed by the RI of the film. Two sequential measurements followed a similar process. Incidents of four different angles equally spaced from $\Phi = 55°$ to $70°$ were sequentially shed on the surface of the samples that were placed on a horizontal stage and the corresponding reflected light was collected by a detector. The original spectroscopic data were then fitted using two types of three-layer models (see the Supplementary materials **Fig. S8** for their detailed comparison).

In both models, a two-layer reference model that consists of a pure silicon substrate beneath a 1.72 nm native oxide layer was applied to represent the silicon substrate. Then a third layer was added on top of the reference model to represent the grown h-BN film. We used a Bruggeman effective-media approximation (EMA) layer that mixes the pure h-BN with voids (n =1). The light spot generated by the ellipsometer has a diameter at a scale of hundreds of microns, which is large enough to bring an average effect for the h-BN film. We also used the pure h-BN part as the third layer. The fitting result of this model gave an unacceptable MSE greater than 600, not a suitable fitting parameter.

We performed single photon emission at room temperature by using a conventional confocal, optical microscope and a Hanbury-Brown and Twiss (HBT) interferometer. The emitted light is collected by an air objective (100x, NA 0.95) and detected by using avalanche photodiodes or a spectrometer. A 532 nm CW laser diode was used to excite the SPEs. A dichroic mirror (cutoff wavelength 532 nm) and a long pass filter (cutoff wavelength 550 nm) allowed suppressing the back-reflected pump beam. The signal was fiber-coupled to either a grating spectrometer (Ocean optics, QE65pro) or avalanche photodiodes (Excelitas) with 30% collection efficiency in the relevant wavelength range, and the detection event was recorded using a time-tagged single photon-counting module (TTM8000).

**Supplementary materials**

Supporting information contains further structural, optical and tribological analysis.

**Acknowledgments**

This work was sponsored by the Army Research Office and was accomplished under Cooperative Agreement Number W911NF-19-2-0269. The views and conclusions contained in this document are those of the authors and should not be interpreted as representing the official policies, either expressed or implied, of the Army Research Office or the U.S. Government. The U.S. Government is authorized to reproduce and distribute reprints for Government purposes notwithstanding any copyright notation herein. T. Li and Y. Zhao are supported as part of ULTRA, an Energy Frontier Research Center funded by the US Department of Energy (DOE), Office of Science, Basic Energy Sciences (BES), under Award No. DE-SC0021230. Manoj Tripathi and Alan Dalton would like to thank strategic development funding from the University of Sussex.

**Author Contributions**

A. B., R. V. & P. M. A. conceptualized the study. A.B., A. B. P., S. A. I., C. L., H. K., X. Z., T. G., M. S. R. S. & J. L. grew and characterized the films. F. L., M. T. & A. D. did the adhesion and friction force characterizations. T. L. & Y. Z. carried out the refractive index measurement. R. M., C. Y. C. & A. L. G. conducted the single photon emission experiment. A. G. B., M. R. N., D. A. R., P. B. S. & T. I. commented on the manuscript. All the authors discussed the results and contributed on the manuscript preparation.

**Conflict of Interest**

The authors declare no conflict of interest.

**Data Availability Statement**

The data that support the findings of this study are available from the corresponding authors upon reasonable request.